\documentclass[pra,twocolumn,superscriptaddress]{revtex4}
\usepackage{amsmath,amsfonts,bbm}

\usepackage{color}

\newcommand{\ket}[1]{\left|#1\right\rangle}
\newcommand{\bra}[1]{\left\langle#1\right|}
\newcommand{\tr}{\mathrm{tr}}
\newcommand{\E}{\mathbb{E}}
\newcommand{\D}{\mathbb{D}}
\newcommand{\A}{\tilde{A}}
\newcommand{\capt}[1]{\section{#1}}

\begin{document}

\title{Matrix Product State and mean field solutions for one-dimensional
systems can be found efficiently}
\author{Norbert Schuch}
\affiliation{Max-Planck-Institut f\"ur Quantenoptik,
Hans-Kopfermann-Str.\ 1, D-85748 Garching, Germany.}
\affiliation{Institute for Quantum Information, California Institute of
Technology, MC 305-16, Pasadena CA 91125, U.S.A.}
\author{J.\ Ignacio Cirac}
\affiliation{Max-Planck-Institut f\"ur Quantenoptik,
Hans-Kopfermann-Str.\ 1, D-85748 Garching, Germany.}

\begin{abstract}
We consider the problem of approximating ground states of one-dimensional
quantum systems within the two most common variational ansatzes, namely
the mean field ansatz and Matrix Product States. We show that both for
mean field and for Matrix Product States of fixed bond dimension, the
optimal solutions can be found in a way which is provably efficient (i.e.,
scales polynomially).  This implies that the corresponding variational
methods can be in principle recast in a way which scales provably
polynomially.  Moreover, our findings imply that ground states of
one-dimensional commuting Hamiltonians can be found efficiently.  
\end{abstract}

\maketitle

\capt{Introduction}%
Characterizing the ground states of quantum spin systems is a highly
challenging task. Different from the situation for classical systems,
where for the very least, the ground state can be described efficiently,
this does generally not hold for quantum systems. Thus, their ground
states are considerably harder to compute: as Kitaev has shown, solving
this problem is likely to be hard even for quantum
computers~\cite{kitaev:book,kitaev:qma}.  Even more surprising, finding
the ground states of quantum systems remains equally hard when we restrict
our
interest to one-dimensional systems~\cite{aharonov:1d-qma}.  This is in
sharp contrast to the case of classical spin systems, which can be
efficiently solved in one dimension, whereas two-dimensional systems are
known to be NP-hard.

Despite these hardness results for ground states of quantum systems,
physical properties of interest can frequently be determined efficiently
using numerical methods. Even a mean field ansatz, which completely
neglects correlations between the particles, may already reproduce many
physical quantities relatively well~\cite{chaikin:condensed-matter}.  In
most cases however, correlations are important and other methods must be
applied.  While in two dimensions, imposing frustration or fermionic
statistics yields Hamiltonians which cannot be assessed well by numerical
methods, one-dimensional systems -- despite the hardness results for 1D
Hamiltonians -- generally turn out to be extremely well simulatable using
a method known as the the Density Matrix Renomalization Group (DMRG)
algorithm~\cite{white:DMRG,schollwoeck:rmp}.  DMRG can be understood as a
variational method over the class of Matrix Product States
(MPS)~\cite{ostlund-rommer,frank:dmrg-mps}: These states can be considered
as a generalization of the mean field product states, but with a limited
amount of correlations. It has been proven that this allows for the
efficient approximation of ground states of gapped local
Hamilonians~\cite{hastings:arealaw}.

The motivation for this work originates in the following observation: On
the one side, DMRG is a highly successful algorithm which finds the
correct minimum in basically all practical cases. On the other hand, it
has been shown that the problem of finding MPS ground states can be
NP-hard~\cite{schuch:mps-gap-np}, as well as certain minimization problems
encountered in the DMRG algorithm~\cite{eisert:DMRG-NP}. The contrast
between these hardness results and the apparent success of mean field and
MPS algorithms raises the question whether it is possible to prove the
efficiency of variational methods over the class of mean-field and Matrix
Product States in full generality for gapped quantum systems, or whether
even studying physical 1D problems is hard, as it is the case in two
dimensions.

In this paper, we settle this question by studying whether (and to which
extent) for a given Hamiltonian it is possible to find the optimal mean
field state or MPS for a fixed bond dimension.  Surprisingly, we find that
the same technique which is used to show that classical spin chains can be
solved efficiently also allows for solving the problem in the case of mean
field theories and MPS.

In particular, we show that it is always possible to find the optimal mean
field solution of a qu-$d$-it chain of length $N$ up to accuracy
$1/\epsilon$ in energy in a time which scales as $(N/\epsilon)^{4d}$.
Concerning Matrix Product States, we find that approximating the optimal
MPS of bond dimension $D$ up to precision $1/\epsilon$ requires a
computation time which scales as as $(N^{2d}/\epsilon)^{6D^2}$, where $D$
is the bond dimension of the MPS. For a fixed bond dimension, the scaling
is thus polynomial both in the length of the chain and the accuracy.

Note that the result for MPS (in particular, the exponential scaling in
$D$) is essentially optimal, since is has been shown that the difficulty
of the problem has to scale exponentially with $D$ (more precisely, it is
NP-hard, where $D$ is polynomial in the problem
size~\cite{schuch:mps-gap-np}). Note also that while the exponential
scaling seems daunting, in practice one is typically interested in
intensive quantities for which a moderate $D$ will
suffice~\cite{frank:faithfully}; moreover, the polynomial scaling of the
algorithm in $N$ typically allows for the efficient evaluation of such
quantities even in the thermodynamic limit~\footnote{
In condensed matter problems, one is often interested in the value of
intensive quantities in the thermodynamic   limit, $N\rightarrow\infty$.
By invoking scaling properties typically found in practical situations
\cite{white:DMRG,schollwoeck:rmp}, one can otain the scaling
of the computation time as a function of the desired precision $\delta$
for quantities like the energy density $e(\infty)=\lim E(N)/N$ as follows:
Generally, the error made in estimating $e(\infty)$
goes as $|e(\infty)-E(N)/N| = \mathrm{poly}(1/N)$.
 Thus, in order to compute $e(\infty)$ with accuracy
$\delta$, we need to take $N=\mathrm{poly}(1/\delta)$.
In order to obtain a $\delta$-approximation of a
ground state by an MPS, we need a bond dimension of at most $D =
\mathrm{poly}(N/\delta)$, whereas for gapped Hamiltonians,
$D=\mathrm{poly}(\log(N/\delta))$ will typically suffice. Combining that
with (\ref{eq:total-scaling}), we find that the computation time needed to
compute an intensive quantity in the thermodynamic limit with precision
$\delta$ scales at most as $(1/\delta)^{\mathrm{poly}(1/\delta)}$, whereas
for gapped systems, a quasi-polynomial scaling
$(1/\delta)^{\mathrm{poly}(\log(1/\delta))}$ can be expected.  In
particular, the latter result shows that for gapped Hamiltonians, we can
generally evaluate intensive quantities in the thermodynamic limit in a
way which scales quasi-polynomially in the desired accuracy. 
}.

Our findings show that variational calculations over both the mean field
ansatz as well as Matrix Product States can be carried out in a
way which is promised to yield the optimal solution, thus resolving the
question as to whether variational algorithms over MPS -- such as DMRG --
can be rephrased in a way which provably succeeds.
 Moreover, while the algorithm might be unpractical for $D$'s of
several $100$ as used in practical DMRG implementations, the algorithm
could prove useful in practice to find the optimal MPS ansatz for a low
$D$, which can then be used to bootstrap DMRG~\cite{hastings:private},
thus helping to avoid local minima.  Finally, our findings also imply that
ground states of one-dimensional commuting Hamitonians can be found
efficiently, as they can be expressed as MPS.

\capt{Classical spin chains}%
We start by explaining how to solve a classical spin chain efficiently.
While this is known, it will help us to clarify the essential ideas of the
proof technique, as used later for mean field and MPS.  Given an open
boundary condition (OBC) Hamiltonian
$h_{k,k+1}(i_k,i_{k+1})$,
$i_k\in\{1,\dots,d\}$, $k=1,\dots,N$, we want to 
find the $i_1,\dots,i_N$ which minimize the energy,
\begin{equation}
\label{eq:class-min}
E=\min_{i_1,\dots,i_N}\left[h_{12}(i_1,i_2)+\cdots+h_{N-1,N}(i_{N-1},i_N)\right]\ .
\end{equation}
To this end, define recursively
\begin{align}
\label{eq:class-iter}
E_1(i_1)&=0 \\
E_k(i_k)&=\min_{i_{k-1}}\left[E_{k-1}(i_{k-1})+h_{k-1,k}(i_{k-1},i_k)\right]
   ,\quad k\ge2.
\nonumber 
\end{align}
Then, the ground state energy is given by $E=\min_{i_N}E_N(i_N)$, 
and the minimization can be carried out efficiently by computing the
$E_k(i_k)$ of Eq.~(\ref{eq:class-iter}) sequentially: The reason is that
in every step, $E_k$ (the minimal energy of the half chain left of $k$) only
depends on the value $i_k$ of the spin at $k$ -- this is the only variable
which we still need to access to minimize the energy of Hamiltonian terms
to come. The computational cost is as follows: for each of the $d$ values of
$i_k$, one has to compute $E_k(i_k)$. Each computation involves the
minimization over $d$ settings of $i_{k-1}$, and the total computational cost
is thus $Nd^2$.  Note that the algorithm not only yields the optimal
energy, but also the corresponding ground state $i_1,\dots,i_N$.

The intuition behind the algorithm is to proceed from left to right
through the chain and at every site minimize the energy of the half chain
left of $k$ as a function of the boundary setting. Here, the ``boundary
setting'' contains all those spins whose value will still influence the
optimal energy of  Hamiltonian terms to the right of $k$; in the case of
the classical system, this is just the spin at the boundary. Then, the
optimization can be carried out sequentially by adding one new interaction
at a step and minimizing the total energy of the left half chain (i.e.,
the previous total energy plus the new coupling) as a function of the new
boundary.  Sloppily speaking, the idea is that while proceeding through
the chain, we have to make choices about the state, and we want to keep
the dependence of the minimal energy of the half-chain on all past choices
which can still influence the future.

This construction contains all important ideas for the mean field as well
as for the MPS setting. For mean field, the boundary condition is again
only the spin on the boundary, which however is now  a continuous degree
of freedom. Thus, we have to show that we can discretize its values
without loosing too much accuracy. For
the case of MPS, the situation seems more involved, since all choices in
the past can influence the future. However, as we will show later, MPS
have exactly the property that the influence of past choices on the future
is fully characterized by what is passed through the bonds, and is thus
bounded.

\capt{Mean field}%
Having understood the method, 
let us now consider the problem of minimizing the energy of a 1D quantum
system $H=\sum_{k=1,\dots,N-1}{H_{k,k+1}}$, $\|H_{k,k+1}\|_\infty\le1$ 
with respect to a mean field ansatz 
$ \ket{\psi}=\bigotimes_{k=1}^N\ket{\psi_k}$,
$\ket{\psi_k}\in\mathbb C^d$
(in the following, all states are normalized),
where we try to minimize
\begin{equation}
\label{eq:min-mf}
E=\min_{\ket{\psi_1},\dots,\ket{\psi_N}}
    \sum_{k=1}^{N-1}\bra{\psi_k,\psi_{k+1}}
	H_{k,k+1}\ket{\psi_k,\psi_{k+1}}\ .
\end{equation}
This minimization again allows for a recursive formulation as in
(\ref{eq:class-iter}). However, since the parameters $\ket{\psi_k}$ are
continuous, and the cost functions $E_k(\ket{\psi_k})$ are non-linear and thus
cannot be solved for exactly, we restrict the minimization to an
$\epsilon$-net, i.e., a discrete set of
$\ket{\psi_k^\alpha}$, $\alpha=1,\dots,\mathcal A$ such
that
\[
\forall \ket{\psi_k}\in\mathbb C^d\;
\exists\alpha: \|\psi_k-\psi_k^\alpha\|_1\le\epsilon
\]
(we use the convention $\phi\equiv\ket{\phi}\bra{\phi}$).
As shown in \cite{hayden:epsilon-nets} (Lemma II.4), 
such a net of size $A\le (5/\epsilon)^{2d}$ exists.  
This reduces the algorithm to the algorithm for classical 1D chains
described above, which will yield the optimal solution in the set of all
product states $\bigotimes \ket{\psi_k^\alpha}$ on the net. 

The proper way to think of the net is as a constraint made on the mean
field ansatz as a whole, rather than of a way to approximate each
recursion step separately (since this would lead to accumulating errors).
Thus, in order
to bound the error made by introducing the net, we just have to bound the
difference between the minimum in the set of product states and the
minimum on the net in Eq.~(\ref{eq:min-mf}).  
To this end, start from the optimal product state
$\bigotimes\ket{\psi_k}$ and replace each state by an $\epsilon$-close
state $\ket{\psi_k^\alpha}$ on the net.
For each term in the Hamiltonian, 
this yields an error of at most
\begin{align*}
\big|\tr[H_{k,k+1}&(\psi_k\otimes\psi_{k+1}-
    \psi_k^{\alpha(k)}\otimes\psi_{k+1}^{\alpha(k+1)})]\big|
    \\
&\le
\|H_{k,k+1}\|_\infty 
\|\psi_k\otimes\psi_{k+1}-
    \psi_k^{\alpha(k)}\otimes\psi_{k+1}^{\alpha(k+1)}\|_1 \\ 
 &\le 2\epsilon\ .
\end{align*}
Thus, the total error from restricting the minimization of the energy 
(\ref{eq:min-mf}) to states in the net
is $2N\epsilon$. For a targeted accuracy $\delta$ in energy,
we thus have to  choose $\epsilon=\delta/2N$, so that the size of
each net will be $\mathcal A=(10N/\delta)^{2d}$. The minimization can be
thus rewritten in the iterative form (\ref{eq:class-iter}) and carried out
in time $d^4\mathcal A^2 N = Nd^4(10N/\delta)^{4d}$.

\capt{Matrix Product States}%
Let us now turn towards matrix product states.
Starting again from a 1D quantum Hamiltonian 
$H=\sum_{k=1}^{N-1}H_{k,k+1}$, $\|H_{k,k+1}\|_\infty\le1$,
we wish to minimize the energy over all Matrix Product 
States~\cite{perez-garcia:mps-reps} 
\[
\ket{\chi(\{A^k\})} = \tr[A^1_{i_1}\cdots A^N_{i_N}]\ket{i_1,\dots,i_N}
\]
of a given bond dimension $D$, $A^k_i\in M_{D\times D}$ (except for
$A^1_i\in M_{1\times D}$, $A^N_i\in M_{D\times 1}$). It is known
(cf.~\cite{perez-garcia:mps-reps}) that every
MPS can be brought to a standard form for which
$\sum_i{A^k_i(A^k_i)^\dagger} = \openone$.
With this gauge, 
$\ket{\chi(\{A^k\})}$ is normalized, and
the energy of a term $H_{k,k+1}$ can be computed as
follows: Define
\begin{align}
\rho_1&=1\ ,\nonumber\\
\label{eq:rho-recursion}
\rho_{k+1}=\mathcal R(A_{k},\rho_{k}) & :=
    \sum_i(A^k_i)^\dagger\rho_{k}{A}^k_i\ .
\end{align}
Then, the energy of $H_{k,k+1}$ is given by
\begin{align}
\label{eq:mps-en-local}
&\mathcal E_k(A^k,A^{k+1},\rho_k)=\\
    &\quad\sum_{a,b,c,d}
    \bra{a,b}H_{k,k+1}\ket{c,d}
	\tr[(A^{k+1}_a)^\dagger(A^k_b)^\dagger\rho_kA^{k}_cA^{k+1}_d]\ .
    \nonumber
\end{align}
The task is to minimize the total energy over the set of MPS,
\begin{equation}
E=\min_{A^1,\dots,A^N}\sum_{k=1}^{N-1} 
    \mathcal E_k(A^k,A^{k+1},\rho_k)\ .
\label{eq:mps-emin-onlyA}
\end{equation}
Note that by virtue of the recursion relation $(\ref{eq:rho-recursion})$,
$\mathcal E_k$ actually depends on \emph{all} $A^l$ with $l\le k+1$, thus
seemingly ruling out the same approach as for the mean field ansatz.

To resolve this, we rewrite the minimization over
all $A_k$ in (\ref{eq:mps-emin-onlyA}) as a series of constrained
minimizations,
\[
\min_{A_1,\dots,A_N}=
\min_{A_{N-1},A_{N},\rho_{N-1}}
\!\cdots\!
\min_{(A_{k-1},\rho_{k-1})\rightarrow\rho_{k}}
\!\cdots\!
\min_{(A_1,\rho_1\equiv1)\rightarrow\rho_2}
\]
where ${(A_{k-1},\rho_{k-1})\rightarrow\rho_{k}}$ denotes the tuples
$(A_{k-1},\rho_{k-1})$ which are compatible with $\rho_{k}$, i.e., 
$\mathcal R(A_{k-1},\rho_{k-1})=\rho_k$. Based on this rephrased form of
the minimization (\ref{eq:mps-emin-onlyA}), we define
\begin{align*}
&E_k(A^{k+1},\rho_{k+1})=\\
&\qquad\min_{(A_{k},\rho_{k})\rightarrow\rho_{k+1}}
\mathcal E_k(A^k,A^{k+1},\rho^k)
+E_{k-1}(A^k,\rho_k)
\end{align*}
(if the minimum is over an empty set, let $E_k=+\infty$), 
with $E_0\equiv0$. Then, we can
sequentially solve for the $E_k$, always only keeping track of them as a
function of $A^k$ and $\rho_k$ at the boundary, and thus solve
the minimization problem (\ref{eq:mps-emin-onlyA}).

Clearly, it will again be necessary to use nets to be able to implement
the optimization efficiently.  We will put a net on both the $A$'s and
the $\rho$'s, and denote elements of the nets by $\tilde A$ and
$\tilde\rho$. Let us define an operation 
$\mathcal N:\rho\mapsto\tilde\rho$ which maps every $\rho$ to the closest
point in the net. We define a netified recursion relation for the
$\tilde\rho$'s, $\tilde{\mathcal R}=\mathcal N\circ\mathcal R$, and 
define all minimizations (in particular the constrained minimizations
$\tilde E_k$) with respect to variables $\tilde A^k$ and $\tilde \rho_k$
in the net, and constraints according to the netified recursion relation
$\tilde{\mathcal R}$~\footnote{Note that we never actually have to
evaluate $\mathcal N$: To compute $\tilde E_k(\tilde A_k,\tilde\rho_k)$,
we loop over all $\tilde A_{k-1}$, $\tilde\rho_{k-1}$ in the net
and check whether $\|\mathcal R(\tilde A_{k-1},\tilde\rho_{k-1})-
\tilde\rho_k\|_1\le\epsilon$.}.

This coarse-grained version of the iterated protocol, which can be carried
out efficiently, will give the optimal solution in an ansatz class which
is defined by the coarse-grained variables \emph{and} the coarse-grained
recursion relation -- note that this is \emph{not} equal to the optimal
solution with respect to the coarse-grained MPS due to the additional
coarse-graining $\mathcal N$ in the the $\rho$'s.

To bound the error made by introducing the nets, we compute how much
the energy of an arbitrary MPS changes due to the coarse-graining of 
the $A$'s and the $\rho$'s. 
To this end, we first bound the difference in energy caused by
coarse-graining the $\rho$'s as compared to the energy of the MPS
described by the same $\tilde A^k$'s, and second, we bound the error in
energy made by coarse-graining the $A^k$'s. By choosing the nets such that
both of these errors become small, we make sure that the $\tilde A^k$'s
found by optimizing the above recursion relations yield an MPS which is
close in energy to the optimal MPS.

Let us first bound the error made by inserting $\mathcal N$.  We will put
an $\epsilon_\rho$-net on the $\rho$'s, i.e.\ for each $\rho$
there is a $\tilde\rho$ in the net with
$\|\rho-\tilde\rho\|_1\le\epsilon_\rho$.  From
(\ref{eq:rho-recursion}), 
\begin{align*}
&\|\tilde\rho_k-\rho_k\|_1=\\
&\qquad=\left\|\mathcal N
    \left(\sum (\A_i^k)^\dagger\tilde\rho_{k-1}\A_i^k\right)-
         \sum (\A_i^k)^\dagger\rho_{k-1}\A_i^k\right\|_1\\
&\qquad=
    \left\|
    \sum (\A_i^k)^\dagger(\tilde\rho_{k-1}-\rho_{k-1})\A_i^k
    +\epsilon_\rho\sigma\right\|_1\quad (\|\sigma\|_1\le 1)\\
&\qquad\le
    \left\|
    \sum \A_i^k(\A_i^k)^\dagger(\tilde\rho_{k-1}-\rho_{k-1})\right\|_1+
    \epsilon_\rho\\
&\qquad\le 
    \|\tilde\rho_{k-1}-\rho_{k-1}\|_1+
    \epsilon_\rho\le\cdots\\
&\qquad\le (k-1)\epsilon_\rho\ .
\end{align*}
To bound the effect of the error in $\rho^k$ on each Hamiltonian term in
Eq.~(\ref{eq:mps-en-local}), note that (\ref{eq:mps-en-local}) can be
rewritten as
\[
\mathcal E_k({\A}^k,{\A}^{k+1},\rho_k)= 
\tr[V (H_{k,k+1}\otimes\openone) V^\dagger\rho_k]
\]
with an isometry
$V_{\alpha,ab\beta}=(\A^{k}_a\A^{k+1}_b)_{\alpha\beta}$, 
$\sum_{a,b,\beta}V_{\alpha,ab\beta}\bar V_{\alpha',ab\beta}=
\delta_{\alpha,\alpha'}$.
Then, 
\begin{align*}
&\|\mathcal E_k(\A^k,\A^{k+1},\tilde\rho_k)-
    \mathcal E_k(\A^k,\A^{k+1},\rho_k)\|_1=\\
&\qquad\qquad= 
    \|\tr[ V (H_{k,k+1}\otimes\openone) V^\dagger
    (\tilde\rho_k-\rho_k) ]\|_1\\
&\qquad\qquad\le
    \|V (H_{k,k+1}\otimes\openone) V^\dagger\|_\infty
    \|\tilde\rho_k-\rho_k\|_1 \\
&\qquad\qquad\le (k-1)\epsilon_\rho\ .
\end{align*}
Thus, the total error in energy due to the net on the $\rho$'s is
\begin{align*}
\delta_\rho &= \sum_{k=1}^{N-1}(k-1)\epsilon_\rho 
    \le \tfrac12 N^2\epsilon_\rho\ .
\end{align*}

Second, we have to bound the error in energy made by replacing the $A^k$
by $\tilde A^k$ chosen from an $\epsilon_A$-net, which approximates the
$A_k$ in operator norm up to $\epsilon_A$. To this end, define
$\Delta^k=\tilde A^k-A^k$ and
$\mathbb E_k(\rho)=\sum_i A^k_i\rho\, (A^k_i)^\dagger$,
$\mathbb D_k(\rho)=\sum_i A^k_i\rho\, (\Delta^k)^\dagger$.
Then, $\E_k(\openone)=\openone$, $\E_k$ is contractive with respect to 
$\|\cdot\|_\infty$, and
\[
\|\D_k(\rho)\|_\infty
\le \sum_i \|A_i^k\|_\infty\|\rho\|_\infty \|\Delta_i^k\|_\infty
\le d \epsilon_A \|\rho\|_\infty
\]
The overlap of the MPS with and without a net is given by
\[
\langle\chi(\{A^k\})|\chi\{\tilde A^k\}\rangle
=((\E_1+\D_1)\circ\cdots\circ(\E_N+\D_N))(\openone)\ .
\]
For the iterated application of $(\E_k+\D_k)$, we use that the deviation
$\sigma_k$ from the identity grows according to
\begin{align*}
\|\sigma_{k-1}\|_\infty 
&= \|(\E_k+\D_k)(\openone+\sigma_k)-\openone\|_\infty\\
&\le
    \|\E_k(\sigma_k)\|_\infty+\|\D_k(\openone)\|_\infty+
	\|\D_k(\sigma_k)\|_\infty\\
&\le (1+d\epsilon_A)\|\sigma_k\|_\infty+d\epsilon_A\ .
\end{align*}
Under the condition that 
$\|\sigma_{N-k}\|_\infty\le 2kd\epsilon_A\le1$, this 
yields
$|\langle\chi(\{A^k\})|\chi\{\tilde A^k\}\rangle|
\ge 1-2Nd\epsilon_A$.
From this, we can derive a bound on the difference in energy,
\begin{align*}
\delta_A&=\big|\tr[H(\chi(\{A^k\})-\chi(\{\tilde A^k\}))]\big|\\
&\le
|H\|_\infty \|\chi(\{A^k\})-\chi(\{\tilde A^k\})\|_1 \\
&\le  2N
  \sqrt{1-\left|\langle\chi(\{A^k\})|\chi\{\tilde A^k\}\rangle\right|^2}\\
&\le 4N^{3/2}\sqrt{d\epsilon_A}\ .
\end{align*}

Let the targeted error in energy now be $\delta$, and choose
$\delta_A=\delta_\rho=\delta/2$. Thus, we will need nets of precision
$\epsilon_\rho = \delta/N^2\quad$ and
$\epsilon_A = \delta^2/64N^3d$,
respectively.
Such nets exist of size $\mathcal A_\rho = (3/\epsilon_\rho)^{D^2}$ and
$\mathcal A_A=(3/\epsilon_A)^{2dD^2}$~(\cite{pisier:convex-bodies}, see
Appendix).  Each evaluation of an energy in (\ref{eq:mps-en-local}) takes
$d^4D^{3}$ elementary steps, and thus the total computational cost is
\begin{equation}
        \label{eq:total-scaling}
Nd^4D^3\left[\frac{
    3^{2d+1}2^{12d}N^{6d+2}d^{2d}
    }{\delta^3}\right]^{2D^2}\ .
\end{equation}

\capt{Commuting Hamiltonians}%
Our results also imply that the ground state of a local Hamiltonian on a
one-dimensional chain with mutually commuting terms can be found
efficiently. This follows straight away from the fact that the ground
state of any commuting Hamiltonian can be expressed as an MPS with bond
dimension $D = d^2$~\cite{wolf:mutual-info-arealaw}. 

Alternatively, one can map the problem of solving any one-dimensional
commuting Hamiltonian to solving a classical 1D chain, as shown by Bravyi
and Vyalyi~\cite{bravyi:comm-ham}, which allows to use the classical 1D
algorithm described at the beginning.  (In fact, they show that every
commuting 2-local Hamiltonian -- which is always the case in 1D -- can be
mapped to a classical problem on the same interaction graph.)

\capt{Periodic boundary conditions}%
Up to now, we have focused on Hamiltonians on OBC 1D chains. Let us now
briefly discuss how to adapt our method to the case of a periodic boundary
condition (PBC) 1D system.

In the case of both classical chains and the mean field ansatz, the PBC
case can be tackled by additionally letting all $E_k$ in
(\ref{eq:class-iter}) depend on $i_1$
($\ket{\psi_1^\alpha})$: the reason is that $i_1$ is also part of the
half-chain boundary, as the value of $i_1$ will influence the optimal
energy when evaluating the Hamiltonian term $h_{N,1}$.

To solve a PBC Hamiltonian with (OBC) MPS, the situation is a bit more
subtle, as the energy of the coupling $H_{N,1}$ does not only
depend on $A^1$, but also on the way in which it is passed through the
chain. Two possibilities to deal with that would either be to evaluate
the $E_k$ not only as a function of $A^k$ and $\rho_k$, but also of the
$\rho_k^{\alpha\beta}$ arising from putting an operator basis
$\ket{\alpha}\bra{\beta}$ at site 1 [i.e., starting the recursion from the
state $\rho_2^{\alpha\beta}=(A^1_\alpha)^\dagger A^1_\beta$]; or to
keep the dependence on $A^1$ and $A^k$, and instead of 
keeping the dependence on $\rho_k$ rather
make $E_k$ depend on the possible channels $\mathcal
R(A^k,\mathcal R(A^{k-1},\cdots,\mathcal R(A^2,\cdot)))$ 
passing $\rho_1$ through the MPS.  A more efficient way to deal with PBC
Hamiltonians is however to fold the chain: This yields an OBC Hamiltonian
on a chain of length $N/2$ which can be solved with the
original algorithm using an OBC MPS with bond dimension $D^2$, which
includes the case of the folded
MPS of dimension $D$. Similarly, the case of PBC MPS with bond
dimension $D$ can be solved by embedding it in an OBC MPS with bond
dimension $D^2$.

\section*{Acknowledgements}
We thank 
B.\ Horstmann, 
A.\ Kay, 
D.\ P\'erez-Garc\'ia, and
K.G.\ Vollbrecht
for helpful discussions and comments. This work has been supported by
the EU (QUEVADIS, SCALA), the German cluster of excellence project MAP,
the Gordon and Betty Moore Foundation through Caltech's Center for the
Physics of Information, and the National Science Foundation under Grant
No. PHY-0803371.

\emph{Note added:} Similar results have been independently obtained by D.\
Aharonov, I.\ Arad, and S.\ Irani~\cite{aai}.

\appendix

\section*{Nets}

We use \cite{pisier:convex-bodies}, Lemma 4.10. It states that in $\mathbb
R^n$ with any norm $|||\cdot|||$, there exists an $\epsilon$-net with
respect to $|||\cdot|||$ for the unit sphere with cardinality at most
$(1+2/\epsilon)^n\le(3/\epsilon)^n$. 

This Theorem can be applied directly to the case of $D\times D$ density
operators which span  a $D^2$-dimensional real vector space. The size of an
$\epsilon_\rho$-net with respect to the trace norm is thus $\mathcal
A_\rho\le (3/\epsilon_\rho)^{D^2}$. To create such a net, one can e.g.\
place a lattice on $\mathbb R^{D^2}$ and use that the trace norm 
is bounded relative to the Euclidean distance.

For the case of the space of isometries $A:\mathbb C^D\rightarrow \mathbb
C^{d}\times\mathbb C^D$, it is necessary to embed them in a vector space,
namely the space of all complex $D\times dD$ matrices. Then, the theorem
states that there is a net of isometries of size at most
$\mathcal A_A\le (3/\epsilon_A)^{2dD^2}$. (Assuming that the net consists
of unitaries imposes no restriction, since the theorem bounds the size of
any maximal set of vectors with distance $\ge\epsilon$.) By utilizing the
Solovay-Kitaev theorem, we can generate such a net
efficiently for any $D$~\cite{kitaev:book,harrow:ba-thesis}.

\end{document}